\documentclass[11pt,a4paper]{article}
\pdfoutput=1
\usepackage{jheppub}
\usepackage{graphicx}
\usepackage{lipsum}
\usepackage{multirow}
\usepackage[all]{hypcap}

\newcommand{\beq}{\begin{equation}}
\newcommand{\eeq}{\end{equation}}

\begin{document}

\title{Resonances from QCD bound states\\and the 750 GeV diphoton excess}

\author[1]{Yevgeny Kats,}
\author[2]{Matthew J. Strassler}

\affiliation[1]{Department of Particle Physics and Astrophysics,
                Weizmann Institute of Science,\\ 
                Rehovot 7610001, Israel}
\affiliation[2]{Department of Physics, Harvard University,
                Cambridge, MA 02138, USA}

\emailAdd{yevgeny.kats@weizmann.ac.il}
\emailAdd{strassler@physics.harvard.edu}

\abstract{Pair production of colored particles is in general accompanied by production of QCD bound states (onia) slightly below the pair-production threshold. Bound state annihilation leads to resonant signals, which in some cases are easier to see than the decays of the pair-produced constituents.  In a previous paper (\href{http://arxiv.org/abs/1204.1119}{arXiv:1204.1119}) we estimated the bound state signals, at leading order and in the Coulomb approximation, for particles with various spins, color representations and electric charges, and used 7 TeV ATLAS and CMS resonance searches to set rough limits.  Here we update our results to include 8 and 13 TeV data.  We find that the recently reported diphoton excesses near $750$~GeV could indeed be due to a bound state of this kind.    A narrow resonance of the correct size could be obtained for a color-triplet scalar with electric charge $-4/3$ and mass near $375$~GeV, if (as a recent lattice computation suggests) the wave function at the origin is somewhat larger than anticipated.  Pair production of this particle could have evaded detection up to now.  Other candidates may include a triplet scalar of charge $5/3$,  a triplet fermion of charge $-4/3$, and perhaps a sextet scalar of charge $-2/3$.}

\maketitle

\section{Introduction}

Any colored particle within the kinematic range of the LHC is guaranteed to be pair produced, with a model-independent minimal rate determined by its color representation, spin and mass. However, whether the particle will be discovered, and how easy it will be to discern its properties, depends very strongly on its decay modes. For example, many efforts have been dedicated to searches for stops (superpartners of the top quark). While in some scenarios stops have been excluded up to above $1$~TeV (e.g.,~\cite{Chatrchyan:2013xsw}), certain regions of the parameter space still allow stops even lighter than $200$~GeV~\cite{Aad:2015pfx}.

For this reason, it is useful to keep in mind that in addition to pair production at $\sqrt{\hat s} > 2m$, any colored particle also has QCD bound states produced at $\sqrt{\hat s}$ slightly below $2m$. When these bound states annihilate, they produce resonant signals in the dijet, diphoton, dilepton, photon+jet, and other channels. Many of these signals are determined in a largely model-independent way by the particle's mass, spin and gauge quantum numbers, and are present regardless of how obscure the usual decays of the particle might be. As we argued in~\cite{Kats:2012ym}, these resonant signals can play an important role in determining the particle's properties. In some scenarios, they can even be a discovery channel, as we will illustrate in the context of the recently reported excesses near $M\sim 750$~GeV in the ATLAS and CMS diphoton resonance searches~\cite{ATLAS-CONF-2015-081,CMS-PAS-EXO-15-004}.

In \cite{Kats:2012ym} we computed the bound state annihilation signals, at leading order and in the Coulomb approximation, for a variety of particles.\footnote{For other studies of bound state phenomenology at the LHC, see~\cite{Drees:1993yr,Drees:1993uw,Chikovani:1996bk,Arik:2002nd,Cheung:2004ad,BouhovaThacker:2004nh,BouhovaThacker:2006pj,Bussey:2006vx,Martin:2008sv,Kim:2008bx,Kauth:2009ud,Kats:2009bv,Younkin:2009zn,Kahawala:2011pc,Barger:2011jt,Kumar:2014bca,Batell:2015zla,Luo:2015yio,Potter:2016psi}.} Here we update our work to include the results of the 8~TeV and the early 13~TeV searches by ATLAS and CMS. This leads to new model-independent bounds on the possible masses of yet-unknown colored particles, as a function of their spin, color representation and electric charge.

The bound state annihilation signals are complementary to signals from pair production.  On the one hand, they apply even to constituents that decay to obscure final states.  On the other, they can be evaded only if the constituents are short-lived, with a width larger than the bound-state annihilation width ($\sim 10^{-5} m$ for a color triplet), which dilutes the resonances.
But a substantial width is usually only possible if the particle has unsuppressed two-body decays (as in the example of toponium~\cite{Fadin:1987wz}, reviewed in~\cite{Kats:2009bv}), which are often relatively easy to search for~\cite{Aad:2015caa,Khachatryan:2015vaa,Aad:2011yh,ATLAS:2012ds,Chatrchyan:2013izb,Khachatryan:2014lpa}.

We will see that in the case  of the 750~GeV excess, it is possible, if the excess turns out to be a narrow resonance, that it is the signal of a QCD bound state of particles $X$,~$\bar X$ that would not otherwise have been discovered up to now.  (This possibility was also suggested earlier in~\cite{Luo:2015yio}, but with different conclusions.)  The most attractive possibilities are that $X$ is a color-triplet scalar of charge $Q=-4/3$ or $5/3$, or a color-triplet fermion with $Q=-4/3$.%
\footnote{Note added: On the day this paper was submitted to the arXiv, the paper~\cite{Han:2016pab} appeared, which studies the fermion scenario and has some overlap with our results for that case.}
The option of a $Q=-2/3$ sextet scalar may also be viable if its dominant decay is to $\geq 4$ jets.

\section{Approximations and Uncertainties}

In a previous paper~\cite{Kats:2012ym}, we derived the properties of the bound state annihilation signals as a function of the constituent particle's mass $m$, spin ($j =0$, $1/2$ or $1$), color representation ($R = \mathbf{3}$, $\mathbf{6}$, $\mathbf{8}$, $\mathbf{10}$, $\mathbf{15}$, \ldots) and electric charge $Q$. We assumed these particles have no sizable couplings apart from the Standard Model gauge interactions; in particular we assume their coupling to the Higgs is small and that all fermions are vector-like.  With the Coulomb approximation for the QCD binding potential, these calculations are straightforward.  Importantly, in evaluating the long-distance part of the matrix elements (via the square of the wave function at the origin) we use $\alpha_s$ evaluated at the scale of the state's typical size, $\sqrt{\langle r^2\rangle}$, while in the short-distance parts of the production and annihilation matrix elements we evaluate $\alpha_s$ at the mass $m$.  
This approximation is expected to be fairly reliable for particles with masses of ${\cal O}(100~{\rm GeV})$ and above, since $\alpha_s \sim 0.1$ and the bound state size, of order $2/(\alpha_s m)$, is much smaller than the confinement scale, $1/\Lambda_{\rm QCD}$.  Uncertainties from varying the scale at which $\alpha_s$ is evaluated are of order $25\%$.  

Our calculation can be improved using a potential that includes the running coupling to some order and the long-distance non-perturbative effects extracted from charmonium and bottomonium, as in \cite{Hagiwara:1990sq,Strassler:1990nw}.   However, other higher-order corrections are important too.  In toponium, an NNLO calculation~\cite{Hoang:2000yr} differs significantly from the NLO result; fortunately, as recently shown~\cite{Beneke:2015kwa} through an N$^3$LO calculation, this order appears sufficient for a stable computation of the $t\bar t$ threshold, which is tamed in part by the top quark's width.  For a more complicated structure with narrow resonances, we are not sure of the situation.  In \cite{Kauth:2009ud} it is shown that for a bound state of color-octet fermions, there is a large positive NNLO correction to the wave function at the origin.

One might expect that at least the uncertainties for color triplets are by now under control, given the level of interest in stoponium and heavier quarkonium.  However, this has recently come into question.  On the one hand, it was shown in~\cite{Martin:2008sv} that the diphoton resonance arising from stoponium of mass 750 GeV,  computed at leading order using the potential approximation suggested in~\cite{Hagiwara:1990sq}, is about 0.06 fb at 13 TeV.  It was then shown in~\cite{Younkin:2009zn} that the NLO corrections (and other small effects)
increase the diphoton rate by 25\%, leaving it a factor of 1.7 or so below the Coulomb approximation used in~\cite{Kats:2012ym}.  But in the past year, the lattice work of~\cite{Kim:2015zqa} claims a wave-function-squared at the origin  3.5--4 times larger than the result derived from the potential model approach of~\cite{Hagiwara:1990sq}.  This makes it {\it larger} than the Coulomb approximation. 

If the lattice claim is even partially correct, it would have an important impact.  Scaling up the cross sections from stoponium to constituents with $|Q|=4/3$, one finds that~\cite{Younkin:2009zn} predicts a diphoton resonance with a 13~TeV cross section of order 1.3~fb, while \cite{Kim:2015zqa} predicts one of order 4.5~fb.  Whether a bound state of $|Q|=4/3$ scalars can explain the diphoton excess is therefore dependent on the resolution of the current discrepancy between the two methods.  Certainly the new lattice calculation, given its important claim, should be confirmed by a second calculation.

Since the above-mentioned theoretical discrepancy is so large, and its consequence so significant, the use of a calculation which claims high precision might be misleading, and so we feel we must remain agnostic.  As we did in~\cite{Kats:2012ym}, here we use only leading-order Coulomb-potential expressions for annihilation rates and parton-level production cross sections, along with the NLO MSTW 2008 parton distribution functions~\cite{Martin:2009iq}, to compute the LHC signals for $\sqrt s = 8$~TeV and $13$~TeV.  We therefore accept large uncertainties on our predictions, which we assume to be a factor of two up or down.  This still allows us to obtain interesting bounds, and a small list of candidates which could explain the diphoton excess.  For these candidates, a much more precise treatment is warranted.

A further uncertainty in our results arises because we extract limits by applying ATLAS and CMS results, each of which is obtained  for a fixed spin and production channel, to other spins and channels. 
Since acceptance and signal shape have some dependence on the spin of the resonance, its intrinsic width, and whether a jet is due to a parton-level gluon or quark, this adds to our uncertainies.  We expect this issue is subleading compared to the uncertainties in our signal predictions.

A final uncertainty of importance, when considering whether a bound state signal could explain the diphoton excess at 750 GeV, concerns the size of the excess itself.  A QCD bound state of a particle with non-extreme quantum numbers is necessarily narrow, while the excess reported by ATLAS~\cite{ATLAS-CONF-2015-081} is quite wide, of order 45 GeV with a large uncertainty.  But the excess at CMS~\cite{CMS-PAS-EXO-15-004} is smaller, with no clear preference for a large width, and there are no large excesses at Run~1~\cite{Aad:2015mna,Khachatryan:2015qba,CMS-PAS-EXO-12-045}.  This suggests that the ATLAS excess may be a real signal combined with a large fluctuation, making the excess appear larger and wider than the underlying physical signal.  If so, a narrow resonance could be the source of the excess, but its signal should be on the lower side, perhaps in the 3~fb range. It should certainly not be as large as 10~fb, as would be naively obtained from the ATLAS Run~2 data alone.  We therefore assume that a narrow resonance needs to have a signal between 3~fb -- 6~fb to be a candidate for explaining the excess.  

\section{Limits on Resonances and the 750~GeV Excess}

\subsection{Dijet resonances}

\begin{figure}[t]
\begin{center}
\includegraphics[width=0.48\textwidth]{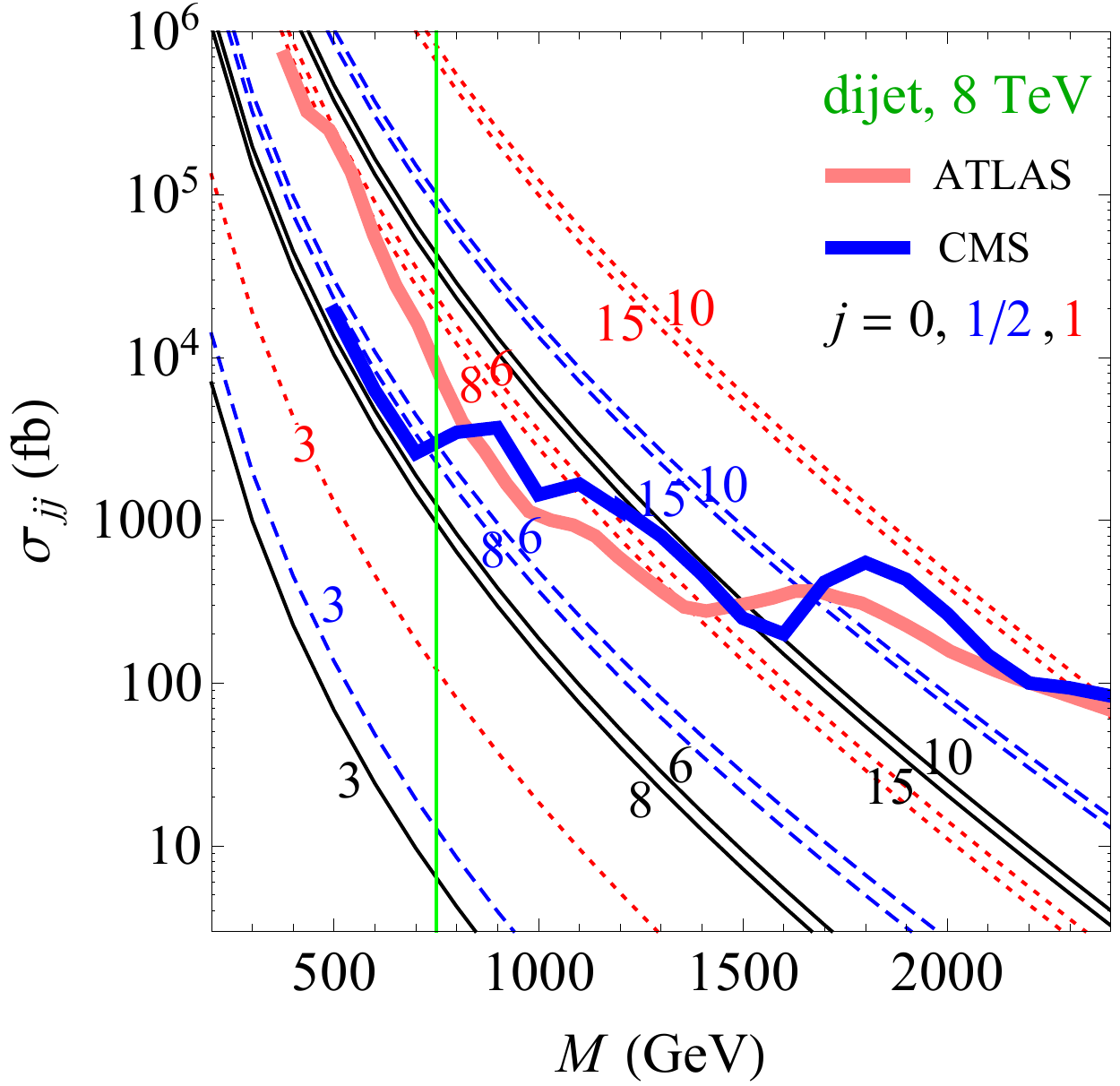}\quad
\includegraphics[width=0.48\textwidth]{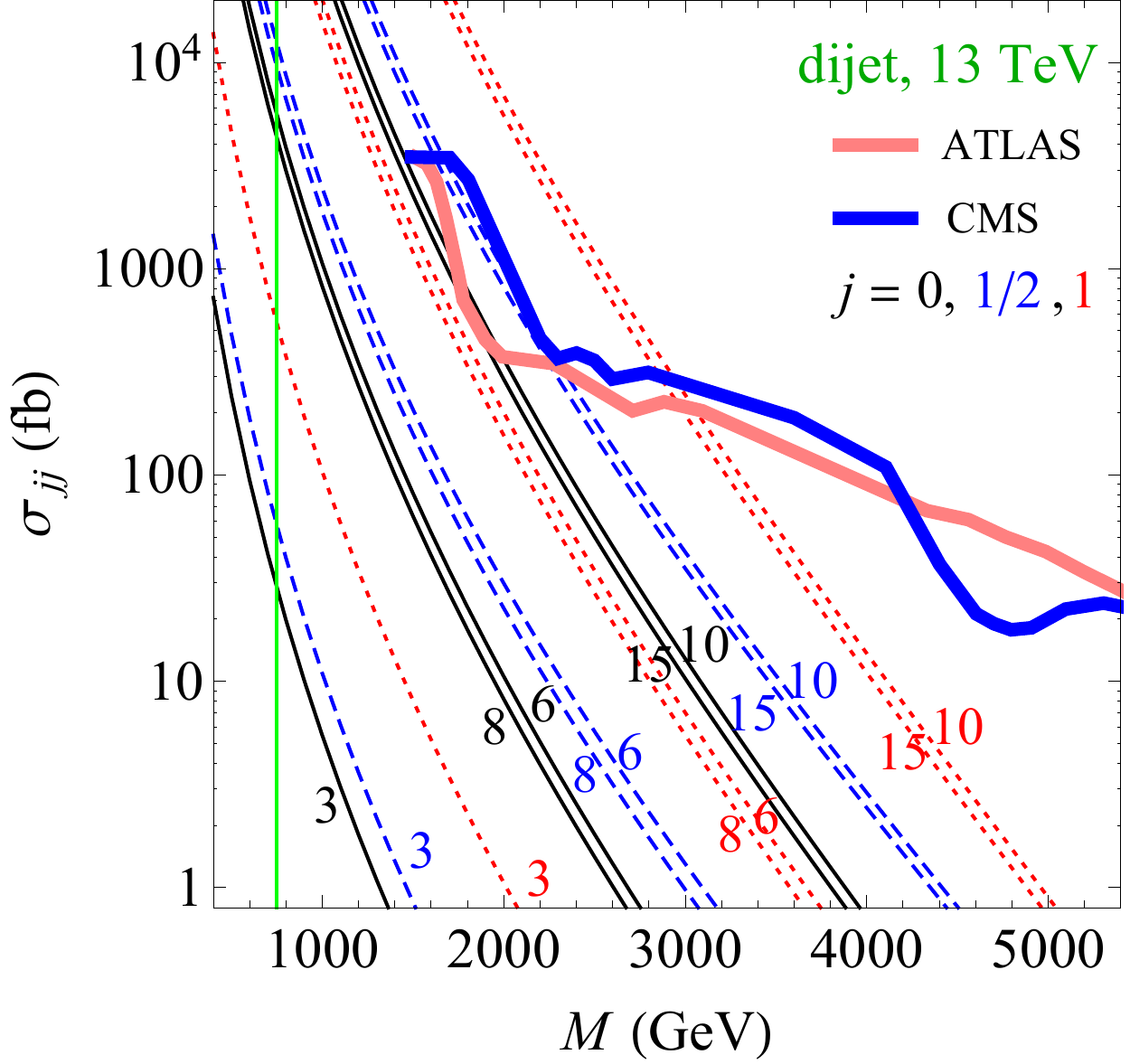}\\
\vspace{-4mm}
\end{center}
\caption{Bound state dijet signals at $\sqrt s = 8$~TeV (left) and $13$~TeV (right) as a function of the bound state mass.  Results are shown for particles with spin $j=0$ (solid black), $1/2$ (dashed blue) and $1$ (dotted red), in color representations $R = \mathbf{3}, \mathbf{6}, \mathbf{8}, \mathbf{10}, \mathbf{15}$, as indicated next to each curve, and are independent of the electric charge. Limits from ATLAS (8~TeV~\cite{Aad:2014aqa}, 13~TeV~\cite{ATLAS:2015nsi}) are shown in thick pink, and CMS (8~TeV~\cite{CMS-PAS-EXO-14-005,Khachatryan:2015sja}, 13~TeV~\cite{Khachatryan:2015dcf}) in thick blue. The green vertical line is at $M = 750~{\rm GeV}$.}
\label{fig-dijet}
\end{figure}

A particle of any spin and in any color representation can form $S$-wave bound states with its antiparticle. These bound states, of various colors, are produced from $gg$ (and from $q\bar q$, in the case of spin-1/2 particles in representations higher than the triplet) via the color gauge interactions of their constituents.  They can similarly annihilate back to $gg$ (or $q\bar q$).  Dijet resonance signals are shown in figure~\ref{fig-dijet}.  The cross sections are plotted as a function of the bound state mass, which is approximately twice the mass of the constituent: $M \approx 2m$.  Also shown are current ATLAS and CMS limits at 8 and 13~TeV.\footnote{For the ATLAS searches~\cite{Aad:2014aqa,ATLAS:2015nsi}, we used the limits provided on Gaussian shapes. Based on a simulation of a scalar $gg$ resonance in~\cite{Kats:2012ym}, we assumed that the peak's mean mass is shifted down by $8.5\%$ relative to the resonance mass, its width is $10\%$, and the acceptance is $30\%$.}  These limits do not constrain color triplets of any spin, and probably allow color-octet and -sextet scalars. Particles in the $\mathbf{10}$ and $\mathbf{15}$ representations are excluded up to $M \sim 1400$~GeV ($m \sim 700$~GeV), so we will not consider them further here. Note the results above are for complex octets; for $Q=0$ an octet can be real, which reduces the dijet signal by 2.

Importantly, these constraints are independent of electric charge, as long as the charge is not large.  For states with large enough charge, the branching fraction of the state to $gg$ may no longer be close to 100\%; decays to electroweak final states may dominate. This happens for $|Q|\gtrsim 2$.

The above results apply to $SU(2)$ singlets.  For $SU(2)$ multiplets of isospin $I$ that have weak couplings to the Higgs boson and so are near-degenerate, the cross section is typically increased by up to a factor of $2I+1$.  This has no serious impact on triplets, but isospin multiplets of octets and sextets are constrained.

\subsection{Diphoton resonances}
\label{subsec:diphoton}

A particle of any spin and in any color representation can form a color-singlet $S$-wave bound state with its antiparticle. If the particle is charged, its spin-0 (and spin-2, if present) bound states, which are produced from $gg$, can annihilate also to $\gamma\gamma$.  Below we consider only particles that can decay to Standard Model particles (and possibly new invisible particles), which restricts $Q$ for a given color representation $R$; for $R = \mathbf{3}$, the allowed values are $Q = -1/3, +2/3, -4/3, +5/3, \ldots$, while for $R=\mathbf{8}$ only integer $Q$ are allowed, etc. In the limit that the annihilation to $gg$ dominates, the $\gamma\gamma$ signals are proportional to $Q^4$.

\begin{figure}[t]
\begin{center}
\includegraphics[width=0.48\textwidth]{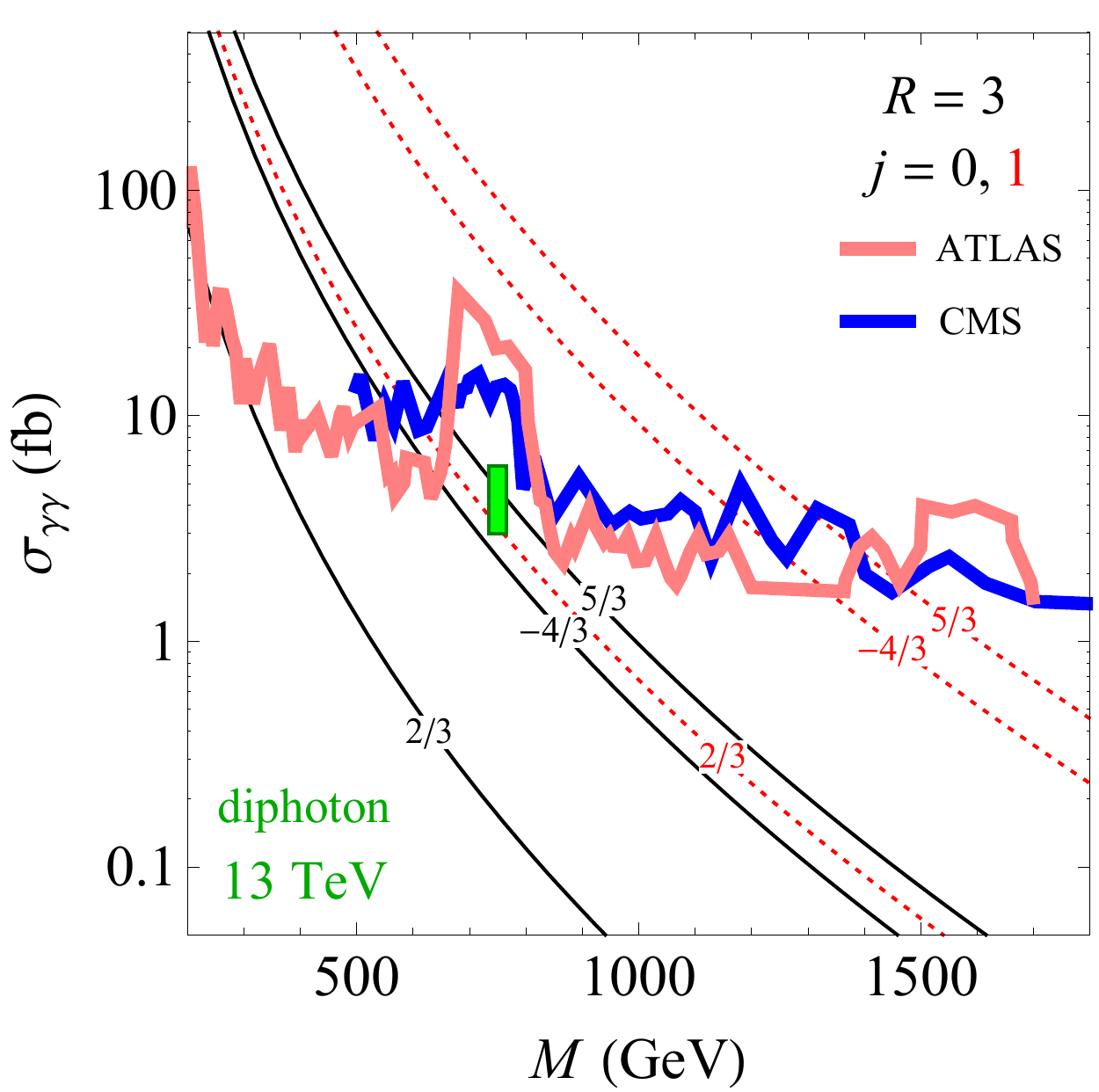}\quad
\includegraphics[width=0.48\textwidth]{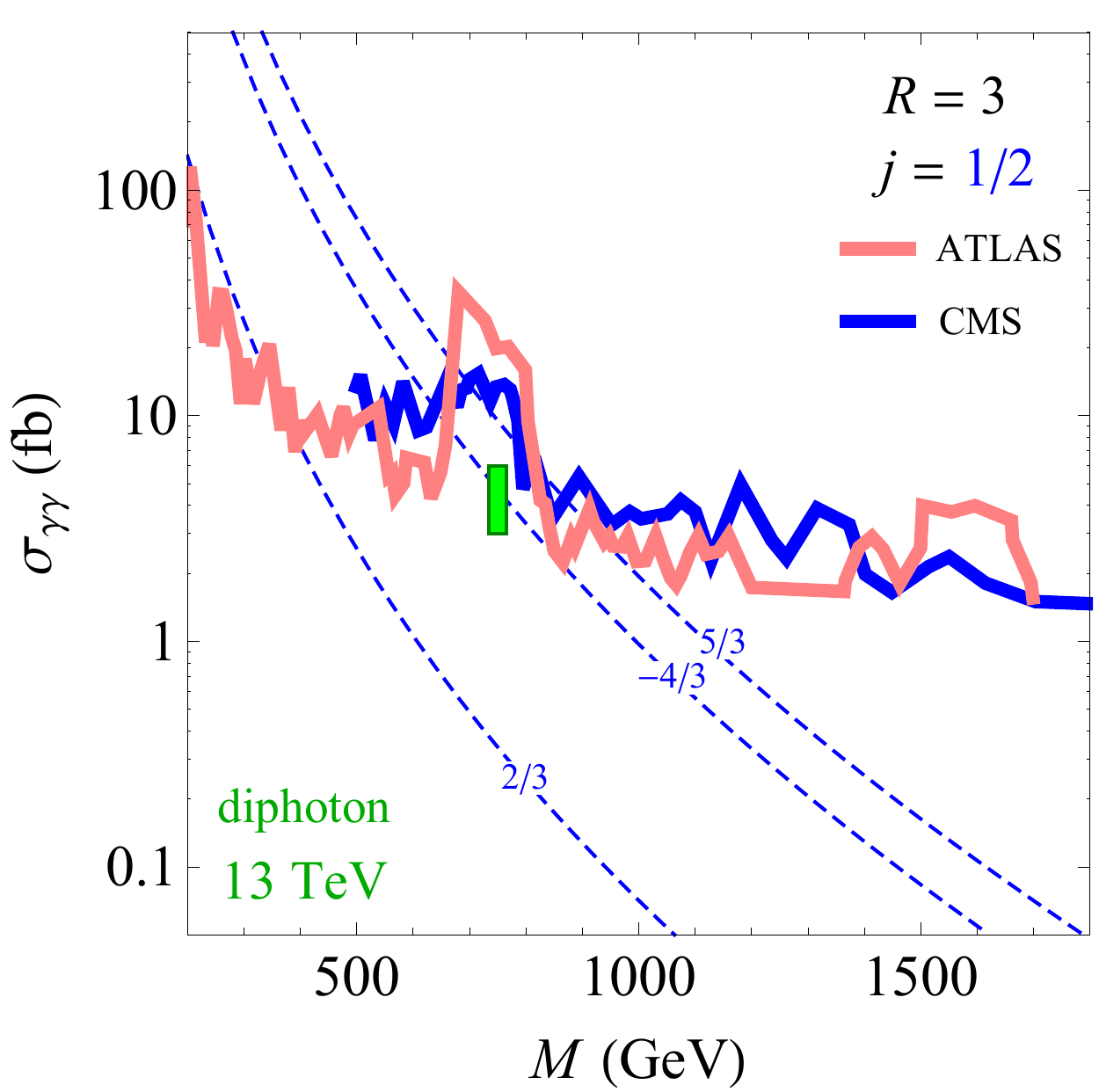}\\
\vspace{4mm}
\includegraphics[width=0.48\textwidth]{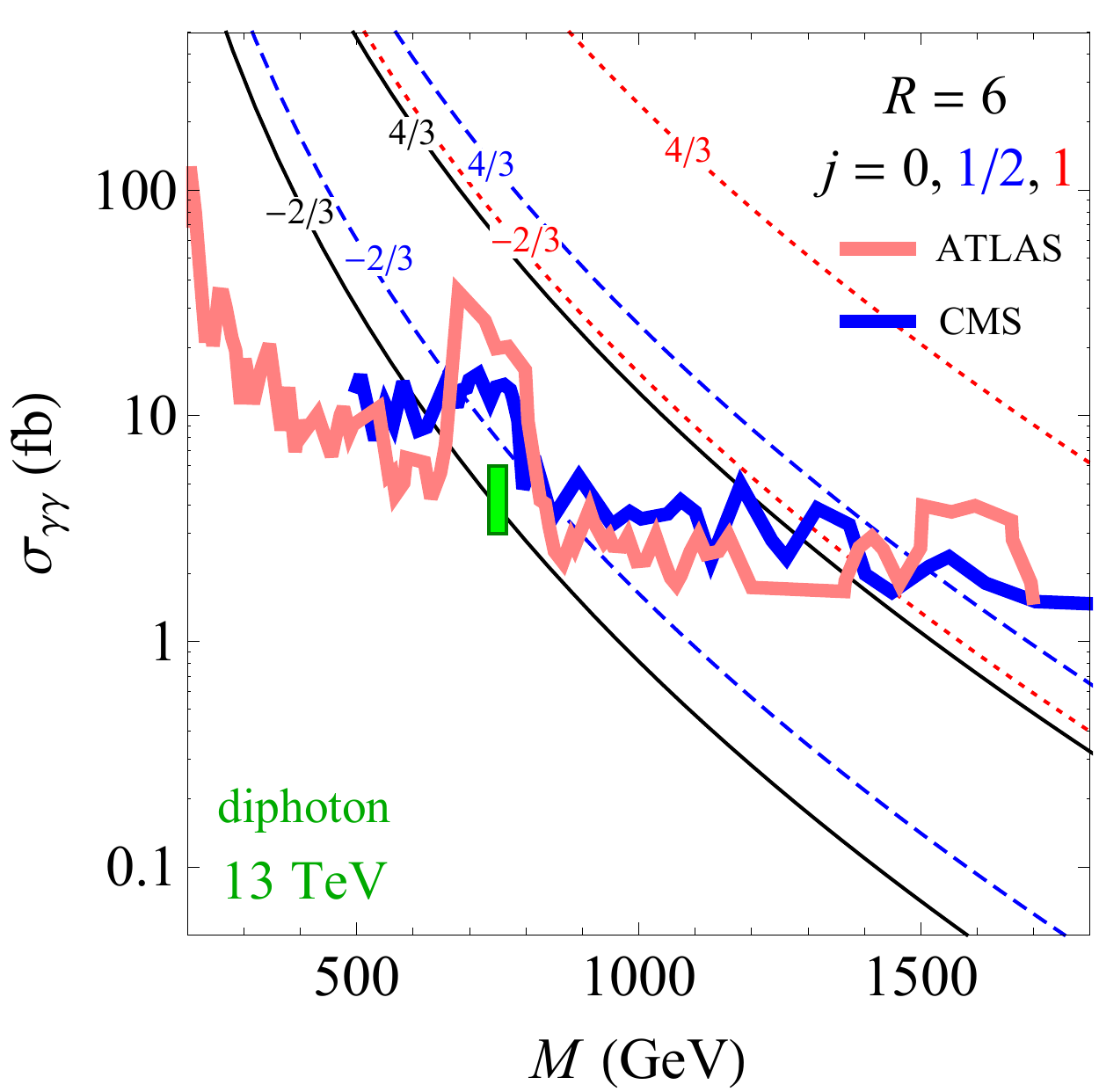}\quad
\includegraphics[width=0.48\textwidth]{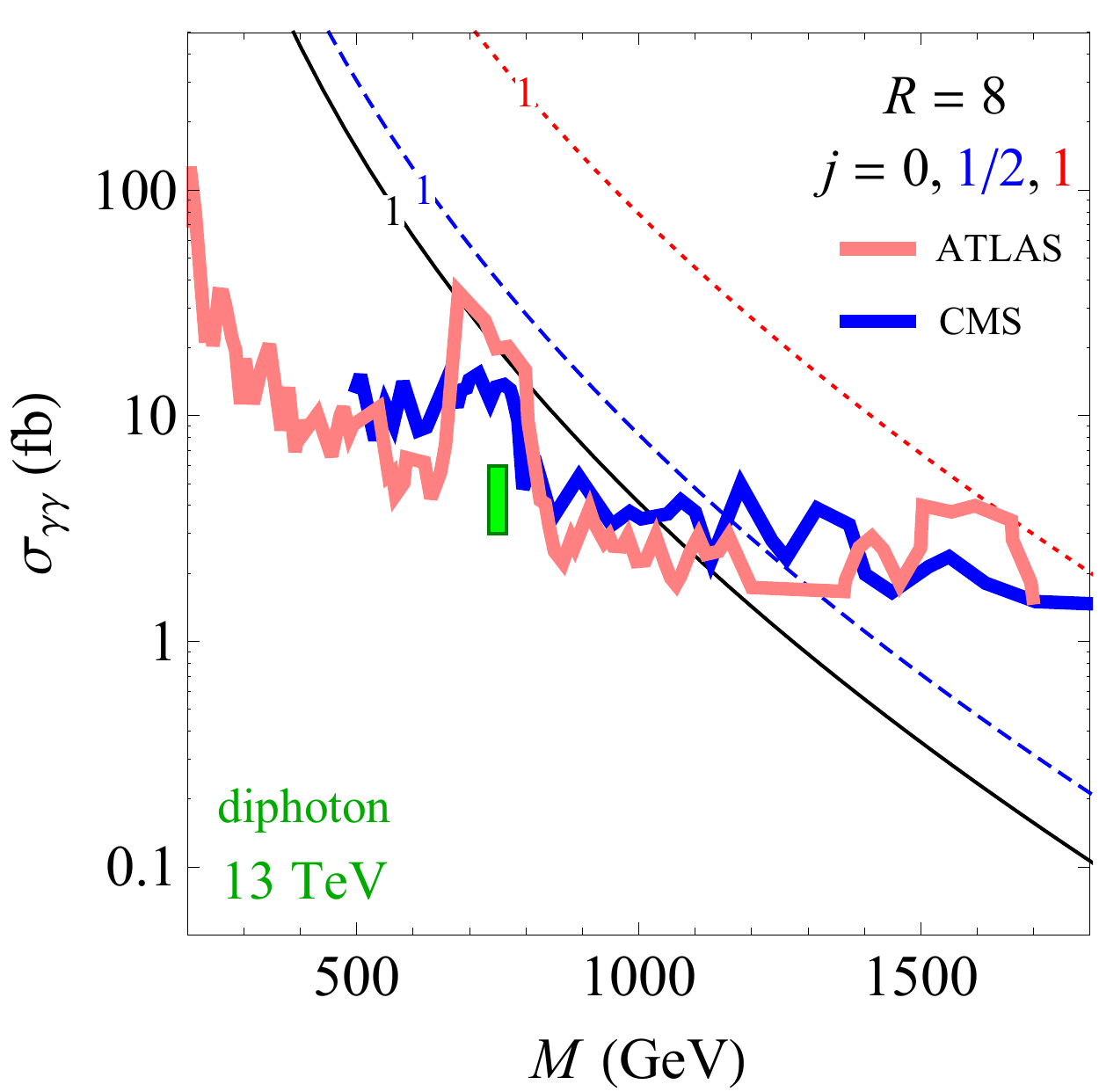}
\vspace{-4mm}
\end{center}
\caption{Bound state diphoton signals at $\sqrt s = 13$~TeV for color triplets (top), sextets (bottom-left) and octets (bottom-right) as a function of the bound state mass. Results are shown for constituents with spin $j=0$ (solid black), $1/2$ (dashed blue) and $1$ (dotted red), for the values of electric charge $Q$ indicated on each curve. Limits are from ATLAS~\cite{ATLAS-CONF-2015-081} (thick pink) and CMS~\cite{CMS-PAS-EXO-15-004} (thick blue). The green rectangle shows the signal size range that can be inferred from the excesses observed in~\cite{ATLAS-CONF-2015-081,CMS-PAS-EXO-15-004} at $M \approx 750$~GeV.}
\label{fig-diphoton}
\end{figure}

Assuming that the total bound state width is dominated by annihilation decays (to $gg$, $\gamma\gamma$, $Z\gamma$ and $ZZ$, for $SU(2)$-singlet particles), the diphoton signals are shown in figure~\ref{fig-diphoton} for color representations $R = \mathbf{3}$, $\mathbf{6}$, $\mathbf{8}$ at $\sqrt s = 13$~TeV. 
Also shown are limits from the ATLAS and CMS diphoton resonance searches.  We indicate the signal size required to explain the excesses at 750~GeV, roughly between $3$ and $6$~fb (see above, and also {\it e.g.},~\cite{Buttazzo:2015txu,Franceschini:2015kwy,Knapen:2015dap,Gupta:2015zzs,Falkowski:2015swt,Buckley:2016mbr}), by a rectangle.

We see, for example, that a stoponium bound state (i.e., $j = 0$, $R = \mathbf{3}$, $Q = 2/3$) would not produce a sufficiently large diphoton signal at $M = 750$~GeV, as has already been noted in~\cite{Luo:2015yio,Gupta:2015zzs}.%
\footnote{Ref.~\cite{Luo:2015yio} proposed that the excess could instead be due to a diquarkonium, a bound state of particles with $j = 0$, $R = \mathbf{6}$, $Q = 4/3$. Our results disagree with the original and revised versions of this paper. According to figure~\ref{fig-diphoton} (bottom-left), the signal of such a bound state would be too large by an order of magnitude.\label{footnote:Luo}} 
On the other hand, a diphoton signal of the right size could arise, for example, from a particle with
\beq
j = 0\,,\quad R = \mathbf{3}\,,\quad Q = -\frac43
\label{candidate-4/3}
\eeq
or
\beq
j = 0\,,\quad R = \mathbf{3}\,,\quad Q = \frac53 \,.
\label{candidate-5/3}
\eeq
The corresponding bound states have $\gamma\gamma$ branching fractions of $8\%$ and $17\%$, resulting in $\gamma\gamma$ signals of $2.3$~fb and $4.8$~fb, respectively, with large uncertainties as discussed above.
For $Q=-4/3$ to explain the diphoton excess it is necessary that the Coulomb prediction is an underestimate, as would be claimed by the recent lattice computation of \cite{Kim:2015zqa}, and not an overestimate as computed in \cite{Martin:2008sv,Younkin:2009zn}.

Another option is
\beq
j = \frac12\,,\quad R = \mathbf{3}\,,\quad Q = -\frac43 \,.
\label{candidate-4/3fermion}
\eeq
Here the $\gamma\gamma$ branching fraction is $8\%$ and the $\gamma\gamma$ signal is $4.7$~fb up to uncertainties.\footnote{Formally, a color triplet of spin 1 and charge 2/3 is also consistent with the data, but to embed such a particle of low mass into a realistic extension of the Standard Model seems daunting.}

Finally we have the option of
\beq
j = 0\,,\quad R = \mathbf{6}\,,\quad Q = -\frac23 \,.
\label{candidate-6}
\eeq
The corresponding bound state has a $\gamma\gamma$ branching fraction of $0.10\%$, resulting in a $\gamma\gamma$ signal of $3.9$~fb.

Each of the above candidates could be part of an $SU(2)$ multiplet, as long as it has the largest electric charge and the mass splittings in the multiplet are too small to allow rapid cascade decays.  Since $\gamma\gamma$ rates go as $Q^4$, the other members of the multiplet do not contribute appreciably to the signal.  However they can contribute to other channels, as we already mentioned above in the dijet case.  There are also changes to the diphoton rates due to $WW$ annihilations and shifts in $ZZ$ and $Z\gamma$ annihilation widths.%
\footnote{For the highest-$|Q|$ particle of an $SU(2)$ multiplet with isospin $I$, the rates are given by
$$\frac{\Gamma_{Z\gamma}}{\Gamma_{\gamma\gamma}} = \frac{2\left(I - |Q|\sin^2\theta_W\right)^2}{Q^2\sin^2\theta_W\cos^2\theta_W}\,,\qquad
\frac{\Gamma_{ZZ}}{\Gamma_{\gamma\gamma}} = \frac{\left(I - |Q|\sin^2\theta_W\right)^4}{Q^4\sin^4\theta_W\cos^4\theta_W}\,,\qquad
\frac{\Gamma_{WW}}{\Gamma_{\gamma\gamma}} = \frac{I^2}{2Q^4\sin^4\theta_W}\,.$$\label{footnote:diboson-dilution}}
For $SU(2)$ doublets this effect is minor for all our candidates because the dominant annihilation channel remains $gg$.

It is also possible to obtain $\gamma\gamma$ signals from bound states of electrically neutral particles, generated via loops involving charged particles. For sufficiently high color representations, the $\gamma\gamma$ signal can be sizable despite the loop suppression. This has been studied for the gluinonium in the MSSM~\cite{Kauth:2009ud,Kahawala:2011pc}. In the context of the 750~GeV excess, the gluinonium signal has been explored in~\cite{Potter:2016psi} and was found insufficient.  As we have noted, higher color representations are excluded by the searches for dijet resonances (figure~\ref{fig-dijet}).

\begin{figure}[t]
\begin{center}
\includegraphics[width=0.485\textwidth]{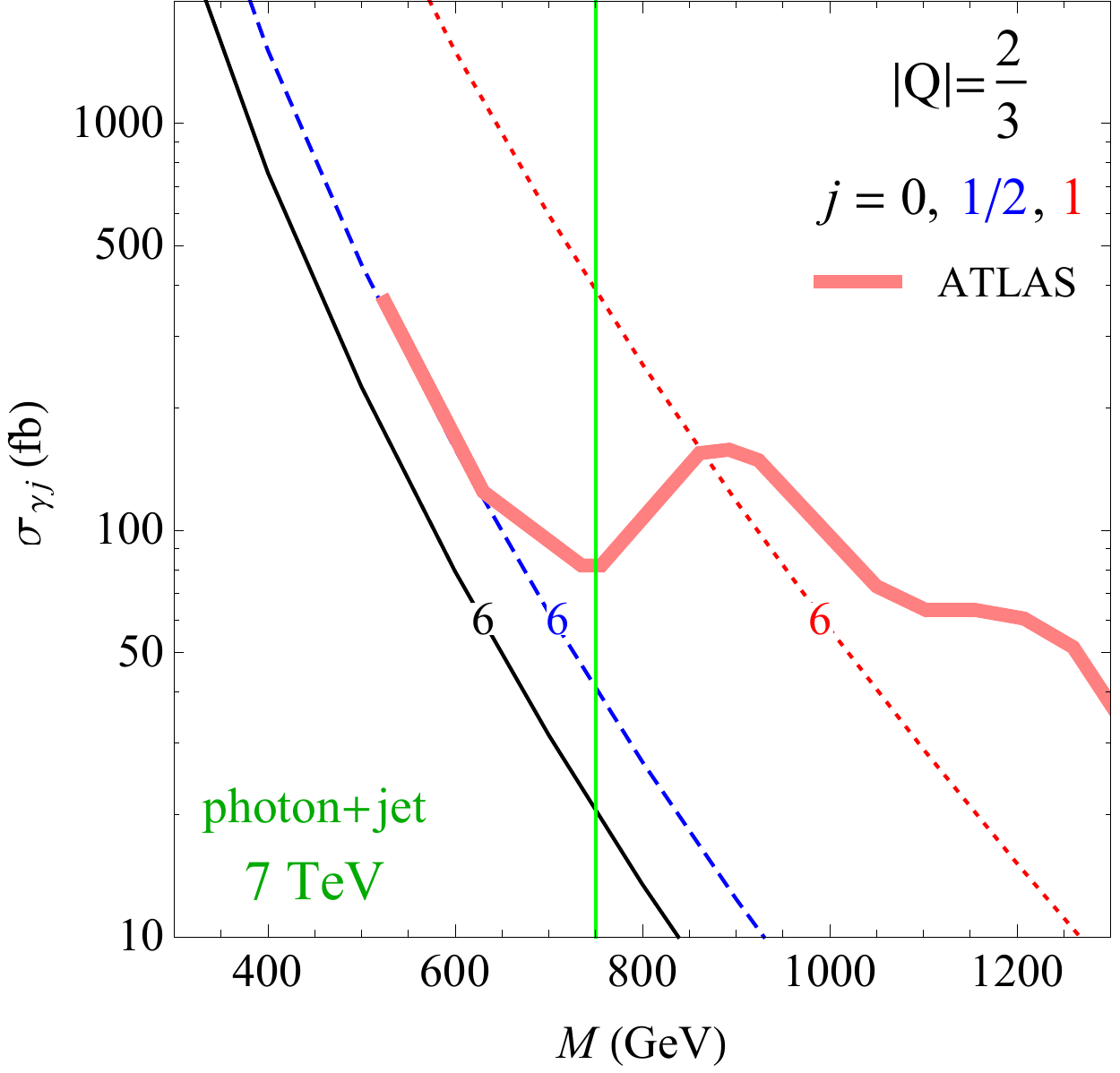}\quad
\includegraphics[width=0.485\textwidth]{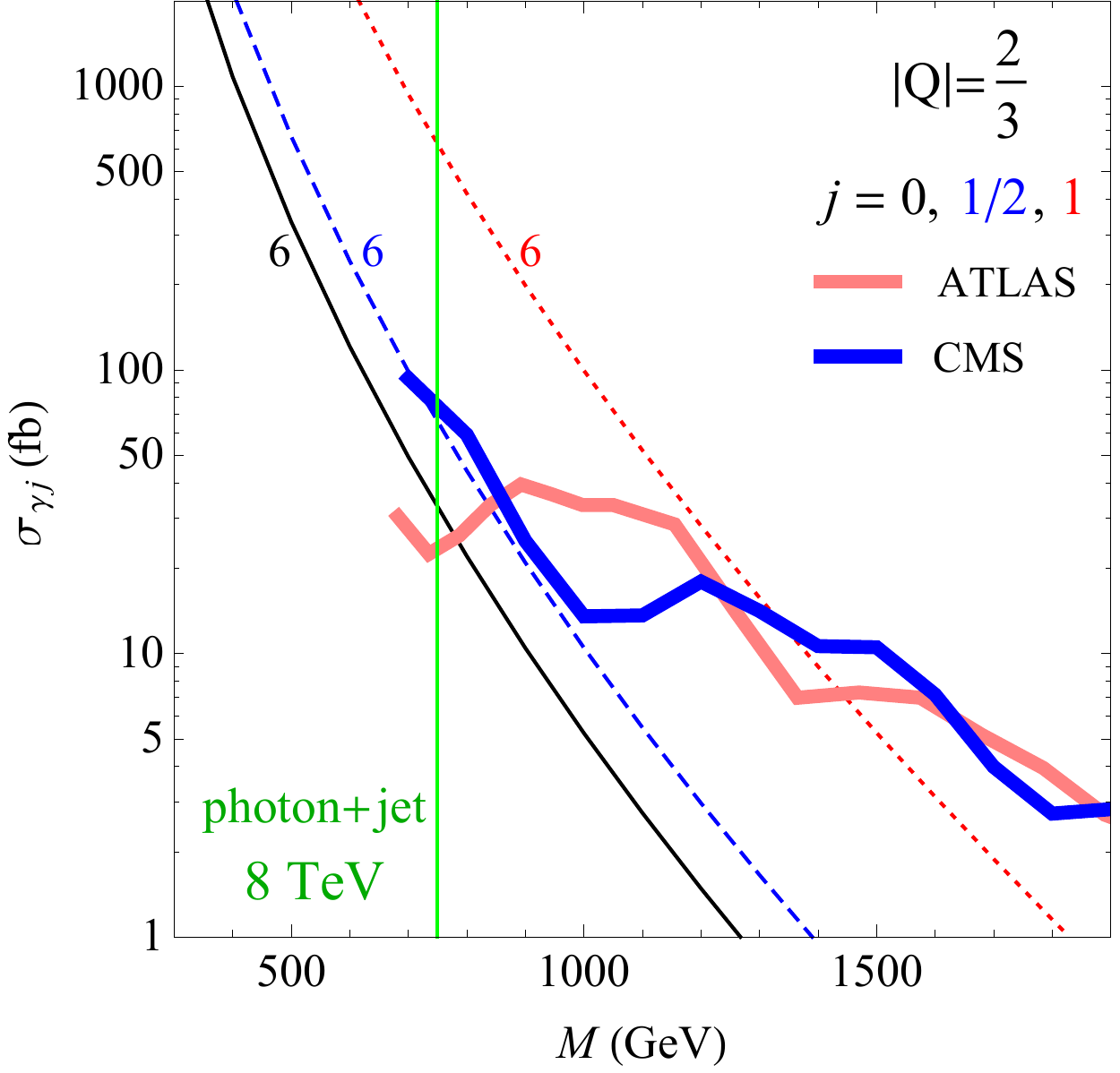}\\
\vspace{4mm}
\includegraphics[width=0.485\textwidth]{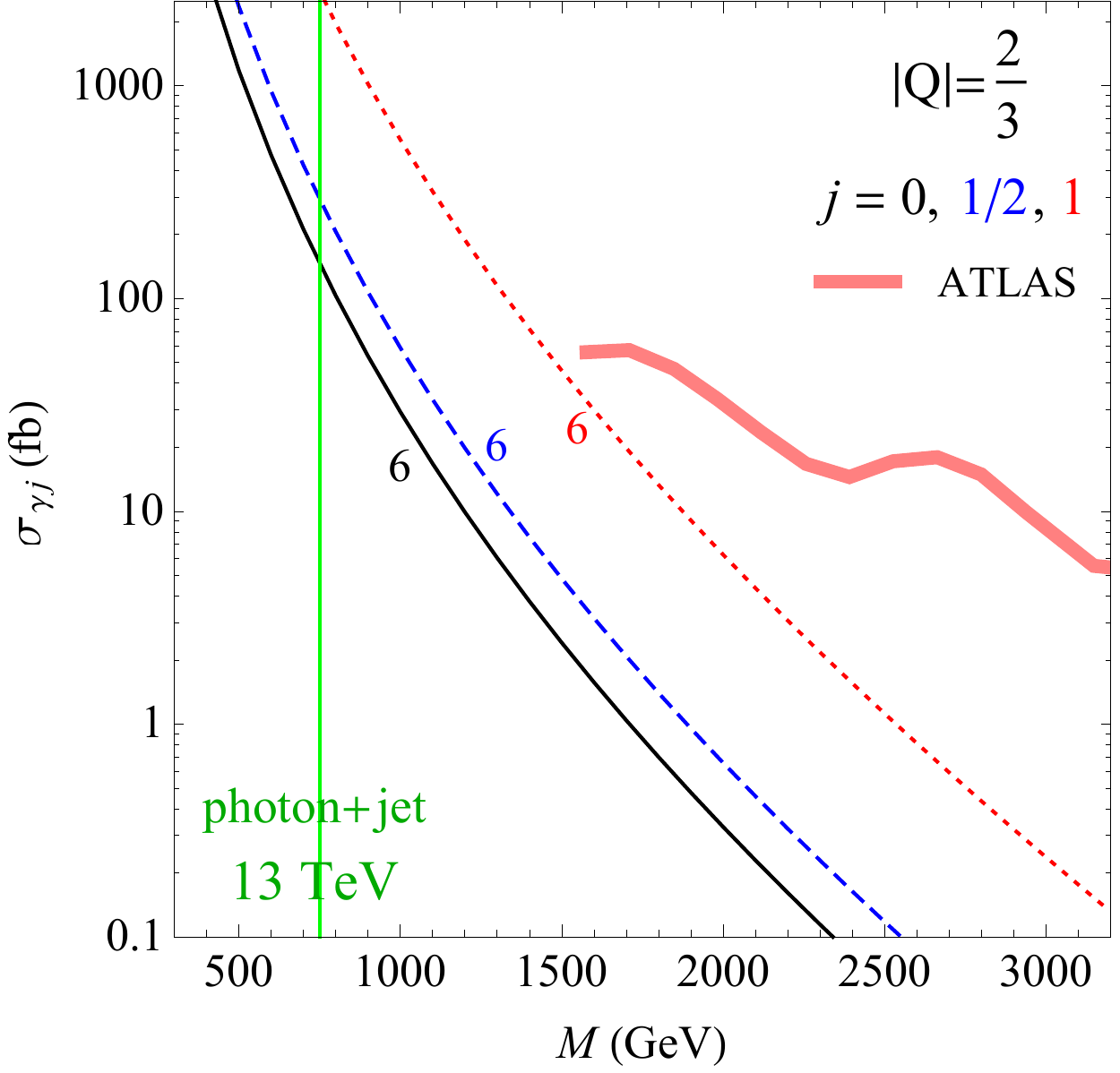}\\
\vspace{-4mm}
\end{center}
\caption{Bound state photon+jet signals for particles with $|Q| = 2/3$ at $\sqrt s = 7$~TeV (top-left), $8$~TeV (top-right) and $13$~TeV (bottom) as a function of the bound state mass. Results are shown for particles with spin $j=0$ (solid black), $1/2$ (dashed blue) and $1$ (dotted red), in the color representation $R = \mathbf{6}$. Limits from ATLAS (7~TeV~\cite{ATLAS:2011ai}, 8~TeV~\cite{Aad:2013cva}, 13~TeV~\cite{Aad:2015ywd}) are shown in thick pink, and CMS (8~TeV~\cite{Khachatryan:2014aka}) in thick blue. The green vertical line is at $M = 750~{\rm GeV}$.}
\label{fig-photon-jet}
\end{figure}

\subsection{Photon+jet, dilepton and other electroweak diboson signals}

Additional constraints arise from searches for photon+jet and dilepton resonances, as we studied in~\cite{Kats:2012ym} and will update here.  There are also potential constraints from $Z\gamma$, $ZZ$ and $WW$ resonances, which we will briefly discuss.  However it appears that none of these can currently eliminate any of the $X$ candidates of eqs.~\eqref{candidate-4/3}-\eqref{candidate-6}.

A $\gamma g$ final state can arise from a color-octet $X\bar X$ bound state for $R=\mathbf{6}$  (as well as for higher colors already excluded by dijet searches), which has mass slightly above the corresponding color-singlet bound state.  Color triplets have no such bound states, while for $R=\mathbf{8}$ the octet states created in $gg$ collisions have the wrong color wavefunction to decay to $\gamma g$.  These photon+jet signals are proportional to $Q^2$.  The possible signals for a sextet with $|Q|=2/3$ are shown\footnote{Note that the formula in~\cite{Kats:2012ym} for the photon+jet signal is too small by a factor of two.  We use the corrected expression here.} in figure~\ref{fig-photon-jet}, along with limits from ATLAS and CMS searches, for $\sqrt s = 7$, $8$, $13$~TeV.\footnote{For the ATLAS searches~\cite{ATLAS:2011ai,Aad:2013cva,Aad:2015ywd}, we used the limits provided on Gaussian shapes. Based on a simulation of a scalar $\gamma g$ resonance in~\cite{Kats:2012ym}, we assumed that the peak's mean mass is shifted down by $5\%$ relative to the resonance mass, its width is $7\%$, and the acceptance is $33\%$.}  We see that the sextet scalar candidate for the diphoton excess is formally excluded by ATLAS~\cite{Aad:2013cva}; however, recall that the signal has considerable uncertainties. It will likely be possible to eliminate or observe this signal in Run~2, as indicated by extrapolating the recent limit of~\cite{Aad:2015ywd}, shown in the lower panel of figure~\ref{fig-photon-jet}, down to $750$~GeV.

\begin{figure}[t]
\begin{center}
\includegraphics[width=0.48\textwidth]{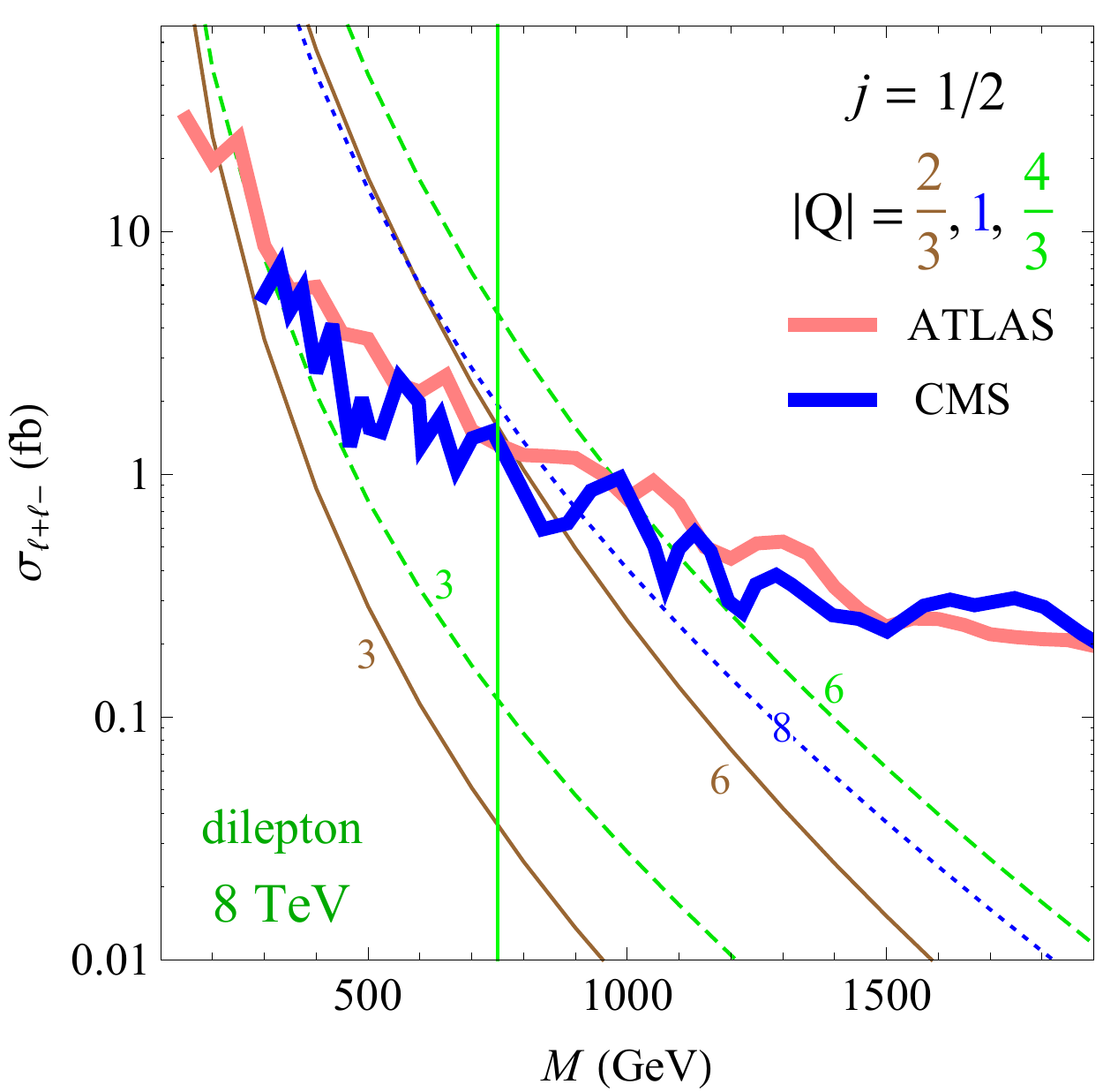}\quad
\includegraphics[width=0.48\textwidth]{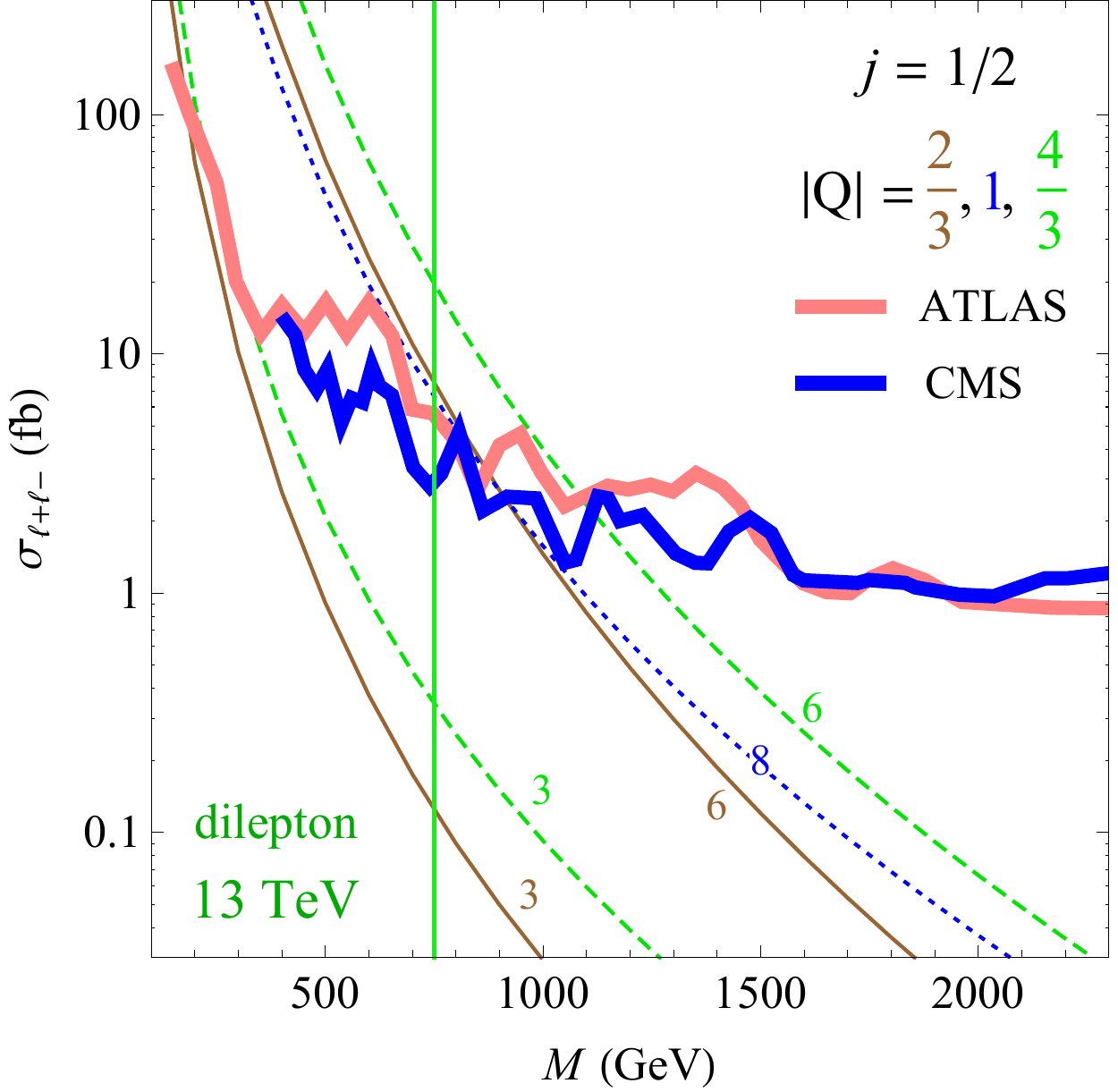}\\
\vspace{-4mm}
\end{center}
\caption{Bound state dilepton signals (for any single flavor of leptons) for spin-1/2 particles with $|Q| = 2/3$ (solid brown), $1$ (dotted blue) and $4/3$ (dashed green) at $\sqrt s = 8$~TeV (left) and $13$~TeV (right) as a function of the bound state mass. Results are shown for particles in color representations $R = \mathbf{3}, \mathbf{6}, \mathbf{8}$, as indicated next to each curve. Limits from ATLAS (8~TeV~\cite{Aad:2014cka}, 13~TeV~\cite{ATLAS-CONF-2015-070}) are shown in thick pink, and CMS (8~TeV~\cite{Khachatryan:2014fba}, 13~TeV~\cite{CMS-PAS-EXO-15-005}) in thick blue. The green vertical line is at $M = 750~{\rm GeV}$.}
\label{fig-dilepton}
\end{figure}

Meanwhile, for spin-$1/2$ fermions, sizable dilepton signals arise when spin-1 color-singlet bound states annihilate via a $\gamma/Z$.  These are absent for scalar constituents.  The current limits on these signals are
shown in figure~\ref{fig-dilepton}. For example, while a color-sextet fermion with $Q=-2/3$ could marginally fit the diphoton excess, it is excluded by the dilepton resonance searches.  On the other hand, the color triplet $Q=-4/3$ fermion candidate of eq.~\eqref{candidate-4/3fermion} is not excluded.

Finally let us comment on  $Z\gamma$, $ZZ$ and $WW$ resonances from these bound states. Sensitivity at 8 TeV to $Z\gamma$ resonances~\cite{Aad:2014fha} was about 10 times weaker than for $\gamma\gamma$~\cite{Aad:2015mna,Khachatryan:2015qba,CMS-PAS-EXO-12-045}, and weaker still for $ZZ$~\cite{Aad:2014xka,Khachatryan:2014gha,Aad:2015kna,CMS-PAS-HIG-14-007} and $WW$~\cite{Aad:2015ufa,Khachatryan:2014gha,Aad:2015agg}.  
It is early days at 13 TeV and there is still no $Z\gamma$ search.  The 13~TeV $ZZ$ and $WW$ searches~\cite{ATLAS-CONF-2015-068,ATLAS-CONF-2015-071,ATLAS-CONF-2015-075,CMS-PAS-EXO-15-002} do not all reach down to 750~GeV, but suggest sensitivity about 50 times weaker in that mass range than the corresponding expected limits from  $\gamma\gamma$ searches~\cite{ATLAS-CONF-2015-081,CMS-PAS-EXO-15-004}.  Of course the sensitivity to the diboson signatures will improve greatly over the next few years as the integrated luminosity increases by two orders of magnitude and the searches are optimized.

The diboson rates, relative to $\gamma\gamma$, are given in table~\ref{tab-diboson}. We consider both the $SU(2)$-singlet case and cases of near-degenerate $SU(2)$ multiplets where we sum the contributions to each diboson final state from all the particles in the multiplet.

For $SU(2)$ singlets, and for $Q=-4/3$ or $5/3$ color-triplets in an $SU(2)$ doublet, the various diboson resonance cross sections from the full multiplet are comparable to that of the  diphoton resonance, which puts these signals well out of near-term reach.   However, for $Q=-2/3$ color-sextets in an
$SU(2)$ doublet,
these resonances can be significantly larger than the diphoton resonance. For a 750~GeV resonance they would perhaps be discovered well before the end of Run~2.

Color-triplet $SU(2)$ triplets require a separate discussion.  Like the isodoublet color sextets, they can have diboson rates larger than for $\gamma\gamma$, and the observed resonances are further enhanced as they receive contributions from all states in the multiplet.  But the $gg$ width of the bound state is much smaller than for a color sextet, and the width to $\gamma Z$, $ZZ$ and $WW$ for the $|Q| = 4/3$ ($5/3$) state is larger by a factor of about $8$ ($3$) than the $\gamma\gamma$ width (see footnote~\ref{footnote:diboson-dilution}). Since the $\gamma\gamma$ branching fraction in the isosinglet case is about $8\%$ ($17\%$), the diboson widths reduce the branching fractions to $gg$ and $\gamma\gamma$ in the isotriplet scenario, affecting the rates shown in figures~\ref{fig-dijet} and \ref{fig-diphoton}. This disfavors the possibility of a $Q=-4/3$ scalar that is part of an isotriplet.

\begin{table}[t]
\begin{center}
\begin{tabular}{c|c|c|c|c}
$I$ & $|Q|$ & $\sigma_{Z\gamma}/\sigma_{\gamma\gamma}$ & $\sigma_{ZZ}/\sigma_{\gamma\gamma}$ & $\sigma_{WW}/\sigma_{\gamma\gamma}$ \\\hline\hline
$0$ & any& 0.63 & 0.10 & 0 \\\hline
\multirow{3}{*}{$1/2$}
 & $2/3$ & 3.7  & 6.3 & 21 \\
 & $4/3$ & 0.32 & 1.1 & 1.4 \\
 & $5/3$ & 0.31 & 0.72 & 0.54 \\\hline
\multirow{2}{*}{$1$}
 & $4/3$ & 3.7  & 6.3 & 14 \\
 & $5/3$ & 1.5  & 3.2 & 6.3 \\
\end{tabular}
\end{center}
\caption{Diboson rates, summed over the $SU(2)$ multiplet, for different choices of the weak isospin $I$ and the highest electric charge in the multiplet, $|Q|$. (For $I = 1$, the numbers depend slightly on the color representation
 and are shown for the color-triplet case.)\label{tab-diboson}}
\end{table}

Thus if any $WW$, $ZZ$ or $Z\gamma$ resonances around 750 GeV were observed soon, it would rule out the isosinglet and isodoublet options, excepting the color sextet isodoublet, though the latter is testable in the photon+jet channel.  If and when such resonances are observed, their relative sizes compared to the diphoton resonance would determine the isospin of the $X$ multiplet and, for nonzero isospin, its hypercharge.

\subsection{Constraints from $X$ pair production}
\label{subsec:decays}

Since model-independent constraints leave several candidates that could produce the observed diphoton excess, we should ask whether other searches for $X$ via its pair production could exclude some or all of them.  As we will now see, firm exclusion is impossible.

\begin{figure}[t]
\begin{center}
\includegraphics[width=0.92\textwidth]{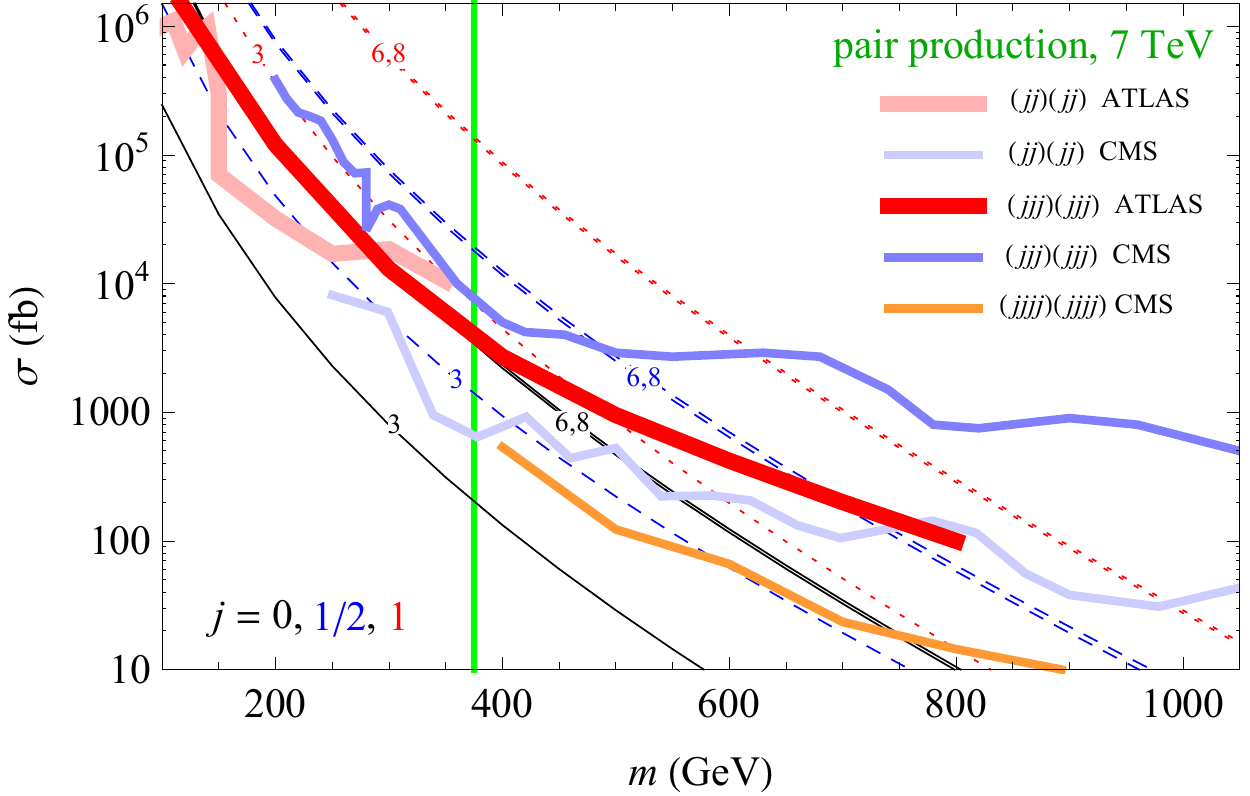}\vspace{6mm}
\includegraphics[width=0.92\textwidth]{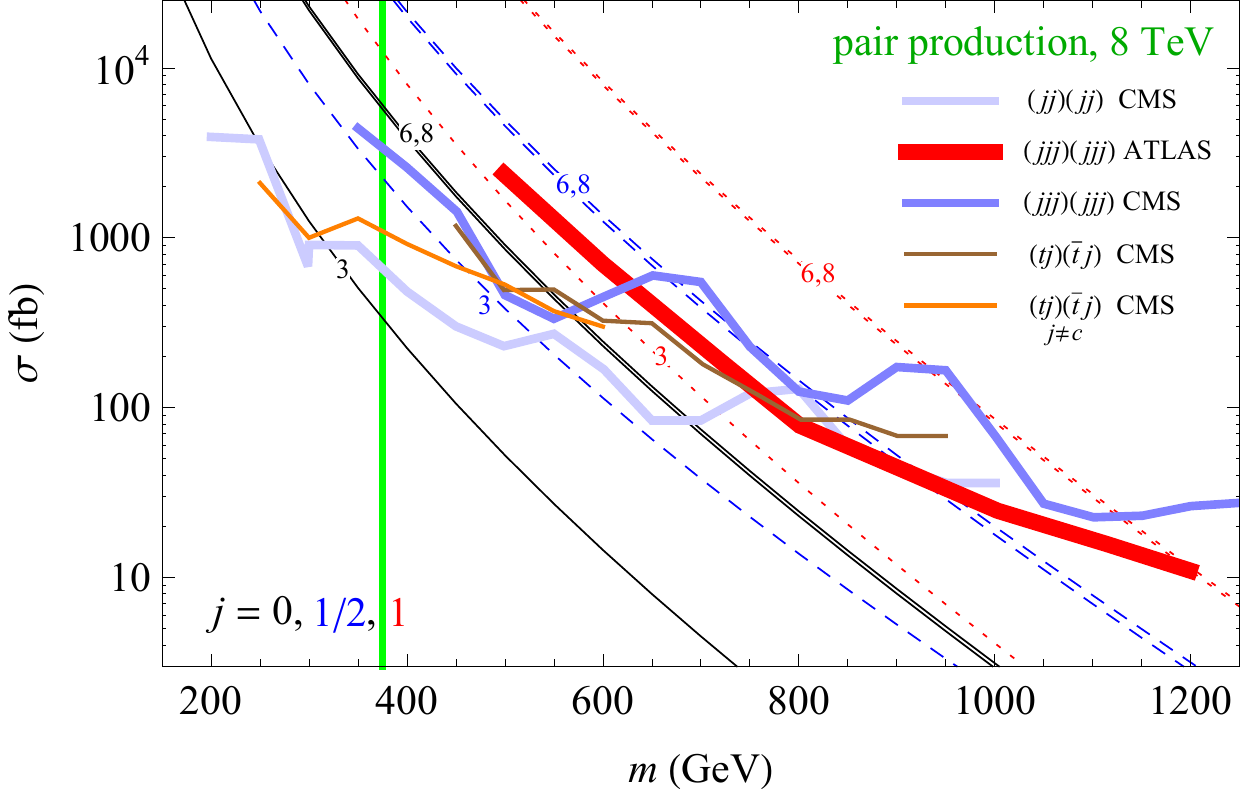}
\vspace{-4mm}
\end{center}
\caption{Limits on pair-produced particles decaying to two, three or four jets, or top+jet, overlaid with their leading-order cross sections (see appendix~C of~\cite{Kats:2012ym}) at $\sqrt s = 7$~TeV (top) and 8~TeV (bottom), for particles with spin $j=0$ (solid black), $1/2$ (dashed blue) and $1$ (dotted red), in color representations $R = \mathbf{3}, \mathbf{6}, \mathbf{8}$, as indicated next to each curve. The limits on $jj$ decays are from refs.~\cite{Aad:2011yh,ATLAS:2012ds}~(7~TeV, ATLAS), \cite{Chatrchyan:2013izb}~(7~TeV, CMS, stop scenario) and \cite{Khachatryan:2014lpa}~(8~TeV, CMS, stop scenario). The limits on $jjj$ decays are from \cite{ATLAS:2012dp}~(7~TeV, ATLAS), \cite{Chatrchyan:2011cj,Chatrchyan:2012uxa}~(7~TeV, CMS), \cite{Aad:2015lea}~(8~TeV, ATLAS) and~\cite{Chatrchyan:2013gia}~(8~TeV, CMS). An example limit on $jjjj$ decays (see text for details) is from~\cite{CMS-PAS-EXO-11-075} (7~TeV, CMS). For $tj$ decays, the limit in brown is from~\cite{Chatrchyan:2013oba} and in orange is from~\cite{CMS-PAS-B2G-12-008} (both are 8~TeV, CMS). The green vertical line indicates $m = 750~{\rm GeV}/2$.}
\vspace{-15mm}
\label{fig-pairs}
\end{figure}

For example, a scalar triplet with $Q=-4/3$ may decay predominantly to pairs of jets through the interaction
\beq
{\cal L} = -\frac{c_{ij}}{2}\,\epsilon_{\alpha\beta\gamma}\,X^{\ast\alpha}\, \overline u_i^\beta\, \overline u_j^\gamma + {\rm h.c.},
\label{int}
\eeq
where $\overline u_i$ is the $i$-th generation $SU(2)$-singlet up-type antiquark field, $\alpha,\beta,\gamma$ are color indices, and \mbox{$c_{ij} = -c_{ji}$} are coupling constants. This dimension-four operator,\footnote{Although $c_{ij}$ violates flavor, it must by assumption be less than $10^{-2}$ (and greater than $10^{-7}$) so that the $X\bar X$ annihilation dominates over $X$ decay (and $X$ decays promptly). It is thus unlikely that $X$ can be observed as a $q q \to \bar X$ resonance.  Moreover, $c_{ij}$ is potentially immune to existing flavor constraints~\cite{Giudice:2011ak}. Such diquarks have appeared, for example, in~\cite{Dorsner:2009mq,Dorsner:2010cu,Dorsner:2011ai,Dorsner:2012nq,Ligeti:2011vt}.} which mediates decays such as
\beq
X \to \bar u\,\bar c \,,
\eeq
is similar to one of the $R$-parity violating (RPV) operators in the MSSM, through which a stop can decay, for instance, as $\tilde{t}_1 \to \bar d\,\bar s$ (see~\cite{Chemtob:2004xr} for a review). The pair-production and decay signatures of the two scenarios are in fact identical, as long as one does not employ charm tagging. The production cross sections are approximately equal since they are dominated by the color gauge interactions. Conveniently, the RPV stop scenario is one of the benchmark models of the CMS searches~\cite{Chatrchyan:2013izb,Khachatryan:2014lpa} and there is currently no exclusion of a top squark at $m \approx 375$~GeV.  By contrast, a charge $-2/3$ sextet decaying to $d_id_j$ is excluded, as seen in the lower panel of figure~\ref{fig-pairs}.  

Depending on the flavor structure of $c_{ij}$ in eq.~\eqref{int}, a color triplet $Q=-4/3$ scalar may also decay as
\beq
X \;\to\; \bar t\,\bar u \;\;\mbox{or}\;\; \bar t\,\bar c \,.
\eeq
These decays seem to be unconstrained as well.\footnote{There is no corresponding mode for a charge $-2/3$ sextet scalar since it cannot decay to up-type quarks.}  The limits from the CMS searches~\cite{Chatrchyan:2013oba,CMS-PAS-B2G-12-008} for pair-produced top-plus-jet resonances are also included in figure~\ref{fig-pairs}. The limit from~\cite{Chatrchyan:2013oba} lies well above the cross section for scalar triplets, and does not even extend down to 375~GeV due to uncontrollable $t\bar t$-plus-jets backgrounds. The search~\cite{CMS-PAS-B2G-12-008} does cover the relevant mass range, but is not sufficiently sensitive either. Moreover, because it requires the light jets to fail a loose $b$-tagging selection, its limit applies only to $X \to \bar t\,\bar u$ decays, while the sensitivity to decays with a charm, $X \to \bar t\,\bar c$, is significantly reduced.

A $Q=-4/3$ color-triplet fermion, if an $SU(2)$ singlet, can decay, for instance, to $d_i(\bar u_j d_k)$ where the quark-antiquark pair are in a color singlet or octet.  If $u_j=u,c$, the resulting three-jet decay is naively disfavored.   As seen in figure~\ref{fig-pairs}, the current limit from CMS~\cite{Chatrchyan:2013gia} is about 3.5~pb, above the leading-order cross section (shown in the figure) but right at the latest precision cross section of $3.5\pm 0.2$~pb~\cite{Czakon:2013tha} (not shown).  At ATLAS, 
the 7 TeV search~\cite{ATLAS:2012dp} does not quite exclude a color-triplet fermion, while the 8~TeV limits~\cite{Aad:2015lea} do not extend to 375~GeV.  However, all these searches assume $X$ decays to three jets with comparable $p_T$. 
If instead the decay occurs in two steps and produces two hard jets and one softer jet, 
then the limits are correspondingly weakened; a similar loophole was pointed out for gluinos in~\cite{Evans:2013jna}.  Meanwhile, searches for RPV gluinos decaying to $tjj$~\cite{Aad:2015lea} may well exclude the case where $u_j=t$, but no explicit limit at 375 GeV is published.   And again there is a loophole if one jet is characteristically softer than the other.  We cannot therefore exclude a $Q=-4/3$ color-triplet fermion, though it may be disfavored unless its decay involves a second (perhaps colorless) on- or off-shell particle with mass near or below $375$~GeV.

An isosinglet color-triplet scalar with $Q=5/3$ cannot decay renormalizably, and typically produces at least four jets if it decays all-hadronically, as in $X\to u_i u_j u_k d_r$ or $\bar d_i \bar d_j \bar d_k u_r$. Something similar applies for a sextet with $Q=-2/3$ if, as required by figure~\ref{fig-pairs}, its diquark decays are suppressed.  For the $X$ to decay promptly, these interactions must be induced by particles whose mass is below, or not too far above, 375 GeV.  Constraints on particles decaying to four jets, for several specific models, were set in~\cite{CMS-PAS-EXO-11-075} based on the 7~TeV dataset. One example of a limit (for a particle with 10\% width decaying via a pair of intermediate particles with mass $m/3$) is included in figure~\ref{fig-pairs}. Color-triplet scalars are clearly allowed. For color sextets, there is not much room. However, considering the subtleties of multi-jet searches (see discussions in~\cite{Evans:2013jna,Asano:2014aka}), it is plausible that some color-sextet scenarios are still allowed. The analogous 8~TeV search~\cite{8jet-thesis} (not shown) is both less sensitive and addresses only much higher masses.

Thus in principle all the candidates mentioned in section~\ref{subsec:diphoton} can have escaped current detection, if their decays are appropriately obscure.  Nor are the cases discussed above the only possibilities.  For example, a $Q=-4/3$ or $5/3$ isodoublet color-triplet scalar could decay, via a soft off-shell $W$, to a slightly lighter $SU(2)$ partner, which in turn decays to two quarks.  This effectively doubles the partner's cross section for $(qq)(\bar q\bar q)$, which is just barely allowed; see figure~\ref{fig-pairs}, and recall the signals shown are leading order.  Another option involves mostly invisible decays of $X$ to soft quark jets plus a new invisible particle, whose mass lies not too far below $375$ GeV. The monojet limits on stop and sbottom~\cite{Aad:2014nra, CMS-PAS-SUS-13-009}, which allow invisibly-decaying triplet scalars with $m\approx 375$ GeV,
and
on a similar scenario with 8 degenerate flavors of light squarks at around $420$~GeV~\cite{Aad:2015iea}, suggest that any invisible or mostly-invisible particle with pair-production cross section $\gg 3$~pb is excluded.  This eliminates the near-invisible option for color sextet scalars and perhaps also for triplet fermions.

In conclusion, we find that existing non-resonance searches do not firmly exclude any of the candidates of eqs.~\eqref{candidate-4/3}-\eqref{candidate-6}, though they do provide important  constraints on the dominant final states in $X$ decays, especially for sextet scalars (triplet fermions) whose simplest decay modes are excluded (disfavored).  In several of the cases we considered, discovery of $X$ pair production is possibly within reach at 13~TeV, though there are a number of possible final states, and the low energies could pose triggering challenges.

\subsection{Width of the 750~GeV resonance}

The 750~GeV excess in ATLAS displays a mild preference for a large width, $\Gamma \approx 45$~GeV~\cite{ATLAS-CONF-2015-081}, though this observation does not have a very large statistical significance.  It also does not receive support from CMS~\cite{CMS-PAS-EXO-15-004} (see also~\cite{Falkowski:2015swt,Buckley:2016mbr}).  Nevertheless it is important to note that the signals discussed here all have a narrow width far below ATLAS and CMS mass resolution.

For the charge $-4/3$ triplet scalar, the annihilation width of the $S$-wave ground state, which makes the dominant contribution to the signal, is $\Gamma \approx 0.005$~GeV, dominated by annihilation to $gg$.  Also, its binding energy is only $E_b \approx -3.4$~GeV, so contributions from excited states at slightly different masses (which altogether increase the cross section by a factor $\sim 1.2$) cannot make the total signal appear wide. The other candidates of eqs.~\eqref{candidate-5/3}-\eqref{candidate-6} still have small widths.  For a sextet the binding is $E_b\sim -20$ GeV, which means that the second resonance, much smaller than the first, could lead to some widening of the currently observed excess, but two separate peaks will be resolved with more statistics.

Thus a confirmation of a single resonance with a large width in future measurements would clearly be inconsistent with the scenario of a single $X$.  Obviously a loophole would remain, in which multiple $X$ particles that are for some reason split by tens of GeV could temporarily mimic a wide resonance.   This would also mean a lower required cross section for each $X$, and therefore lower electric charges, spins and color representations would be preferred.  Constraints from multi-jet production (figure~\ref{fig-pairs}) would then become significant, though still model-dependent.

It is interesting to note that if the bound-state annihilation signals are absent due to overly rapid constituent decays, there will still be a broad feature in the diphoton spectrum above the pair-production threshold, due to loops of $X$. The possibility that this feature is responsible for the $750$~GeV excess has been studied in~\cite{Chway:2015lzg}.

\section{Summary and Discussion}

We have updated our earlier work~\cite{Kats:2012ym} and
determined the current constraints on colored particles that follow from the non-observation of their near-threshold QCD resonances.  We have studied diphoton, photon+jet, dijet and dilepton channels, using the latest ATLAS and CMS searches. Our results for $\sqrt s = 13$~TeV will be useful for interpreting any resonant excesses that might appear in these channels in the course of Run~2.

We also find that the 750~GeV diphoton excess, if it represents a real signal, might be explained by supplementing the Standard Model with a single colored particle $X$ with $m \approx 375$~GeV. 
A narrow $\gamma\gamma$ resonance then arises from the annihilation of an $(X\bar X)$ bound state near the $X\bar X$ pair-production threshold:
\beq
gg \to (X\bar X) \to \gamma\gamma \,.
\eeq
(This was also proposed in~\cite{Luo:2015yio}, but our conclusions differ; see footnote~\ref{footnote:Luo}.)
We have identified a number of candidates for the particle $X$ that would give a signal of the right size, within large theoretical uncertainties.
We saw that in each case
$X$ can potentially decay in a fashion that leaves its pair production hidden from all existing LHC searches.  Unsurprisingly this is most easily accomplished for color-triplet scalars, which have the lowest pair-production cross section.

Among the various candidates listed in section~\ref{subsec:diphoton}, a triplet scalar with $Q=-4/3$ decaying to a pair of up-type antiquarks is perhaps the most theoretically compelling option, as this decay proceeds through a dimension-four operator and does not involve any additional unknown particles. This scenario motivates searches for 375~GeV pair-produced particles, with each decaying to a light jet and a charm jet (e.g., along the lines of the RPV stop searches~\cite{Chatrchyan:2013izb,Khachatryan:2014lpa,Aad:2016kww}), or a top quark and a charm or light jet (similar to the excited top or RPV sbottom signature~\cite{Chatrchyan:2013oba,CMS-PAS-B2G-12-008}).  We see no especially well-motivated search for the other candidates, though they might show up in various existing analyses. All these measurements have challenging backgrounds.

Resonant signals due to annihilation channels other than $\gamma\gamma$ are thus essential to search for.  A resonant dijet ($gg$) signal of roughly $6$~pb at $\sqrt s = 13$~TeV is expected for a sextet scalar and may be easily observable if prescaled or scouting-like triggers~\cite{CMS-PAS-EXO-14-005,Aad:2014aqa} allow. The dijet resonance will be hard to observe  for a triplet scalar or fermion, where it is of order $30$ or $60$~fb respectively.  The sextet's photon+jet resonance should be observable soon; recall it was borderline excluded at Run~1 (figure~\ref{fig-photon-jet}). For the $Q=-4/3$ color-triplet fermion, a dilepton signal of order 0.4~fb  (per lepton flavor) lies in reach, though only with hundreds of inverse fb of luminosity.  Decays to $Z\gamma$, $ZZ$ and $WW$ are small for our candidates if they are $SU(2)$ singlets or doublets, and will not be observed soon, with the exception of an isodoublet $Q=-2/3$ color sextet for which they could be detectable in Run 2. The members of an isospin multiplet will also combine to give  larger signals in both resonance searches and pair-production searches.

Clearly precise calculations of the cross sections of possible diphoton resonances, and in particular a resolution of current discrepancies on which the viability of the $Q=-4/3$ triplet scalar depends, are sorely needed.  If a narrow diphoton resonance is confirmed but no other signal (such as $X$ pair production or another resonant channel) is observed, comparison of the observed diphoton resonance cross section to a precise computation may be the only way to determine whether a bound state of triplet scalars with $Q=-4/3$ or with $Q=5/3$ fits the data.

Let us conclude by noting that the scenario of a QCD bound state is simpler than many models that assume an elementary scalar at 750~GeV. The coupling of an elementary scalar to photons via loops of Standard Model particles alone is never sufficient to account for the signal, so one has to posit the existence of one or more additional charged particles with large Yukawa couplings to the scalar, and/or large electric charges, and/or large multiplicity (see, e.g.,~\cite{Angelescu:2015uiz,Franceschini:2015kwy,Buttazzo:2015txu,Knapen:2015dap,Gupta:2015zzs,Fichet:2015vvy,Agrawal:2015dbf,Aloni:2015mxa,Altmannshofer:2015xfo,Gu:2015lxj}). Our scenario may also be contrasted with many explanations with composite states proposed so far, by noting that no new strong dynamics (e.g.,~\cite{Harigaya:2015ezk,Nakai:2015ptz,Buttazzo:2015txu,Franceschini:2015kwy,Low:2015qep,Bellazzini:2015nxw,Belyaev:2015hgo}) or new confining forces (e.g.,~\cite{Curtin:2015jcv,Agrawal:2015dbf,Craig:2015lra}) are involved in our setup.  In the scenario proposed here, a single particle $X$ near $375$~GeV, pair produced via QCD, weakly bound due to QCD alone, and annihilating via QED, produces a narrow resonance that could perhaps generate the observed excess.  For $X$ a $Q=-4/3$ color-triplet scalar, no other new particles are required and the entire effective theory could be renormalizable.

\acknowledgments

We thank Roberto Franceschini, Elina Fuchs, Diptimoy Ghosh, Rick Gupta, Hyung Do Kim, Steve Martin, Gilad Perez, Michael Peskin, Martin Savage, Matthias Schlaffer and Emmanuel Stamou for useful discussions and/or comments on the draft. 

\bibliographystyle{utphys}
\bibliography{bs-750}

\end{document}